\documentclass[onecolumn,authoryear]{els-mrw} 
\usepackage{amsmath,amssymb,amsfonts,amsthm,makeidx,graphicx}

\usepackage{txfonts}
\usepackage{helvet}

\usepackage{newtxtext}

\usepackage[colorlinks=true, pdfstartview=FitV, linkcolor=magenta,citecolor=blue, urlcolor=magenta,
bookmarks=true, bookmarksnumbered=true, breaklinks]{hyperref}

\newcommand{\beq}{\begin{equation}}
\newcommand{\eeq}{\end{equation}}
\newcommand{\Slash}[1]{{\ooalign{\hfil/\hfil\crcr$#1$}}}
\newcommand{\tr}{{\rm tr}}

\newcommand{\Nc}{N_{\rm c}}
\newcommand{\Nf}{N_{\rm f}}

\newcommand{\lqcd}{\Lambda_{\rm QCD}}
\newcommand{\vp}{\vec{p}}

\newcommand{\vk}{\vec{k}}

\newcommand{\la}{\langle}
\newcommand{\ra}{\rangle}

\newcommand{\calL}{\mathcal{L}}

\newcommand{\rmd}{\mathrm{d}}
\newcommand{\rmi}{\mathrm{i}}
\newcommand{\rme}{\mathrm{e}}

\begin{document}

\chapter{QCD-Like Theories with Different Color Numbers}\label{chap1}

\author[1]{Toru Kojo}%

\address[1]{\orgname{Theory Center}, \orgdiv{IPNS}, 
\orgaddress{High Energy Accelerator Research Organization (KEK), 1-1 Oho, Tsukuba, Ibaraki, 305-0801, Japan}}

\articletag{Chapter Article tagline: update of previous edition,, reprint..}

\maketitle

\begin{glossary}[Glossary]
\textbf{Hadrons}
composite color-singlet states of quarks and gluons bound by the strong interaction.

\textbf{Nuclear forces}
effective interactions between baryons arising from underlying quark and gluon dynamics.

\textbf{Sign problem}
the difficulty in Monte Carlo simulations of QCD at finite baryon density,
caused by the complex fermion determinant.

\textbf{Quark--hadron crossover}
a smooth transition between hadronic and quark matter without a genuine thermodynamic phase transition.

\textbf{Quarkyonic matter}
a hypothetical phase of QCD characterized by a quark Fermi sea coexisting with confining gluodynamics.

\textbf{Chiral Perturbation Theory (ChPT)}
a low-energy effective field theory of QCD based on the spontaneous breaking of chiral symmetry.
\end{glossary}

\begin{glossary}[Nomenclature]
\begin{tabular}{@{}lp{34pc}@{}}
$\Nc$ & Number of colors \\
$\Nf$ & Number of flavors \\
$\mu_q$ & Quark chemical potential \\
$\mu_B$ & Baryon chemical potential ($=\Nc \mu_q$) \\
$n_0$ & nuclear saturation density $ \simeq 0.16\,{\rm fm}^{-3} \sim$ nucleon density in typical nuclei\\
EOS & Equations of state \\
NG bosons & Nambu-Goldstone bosons\\
QC$_2$D & Two-color QCD \\
QCD$_I$ & QCD at finite isospin but zero baryon density \\
BEC & Bose-Einstein condensation \\
BCS & Bardeen-Cooper-Schrieffer \\
pQCD & Perturbative QCD \\
\end{tabular}
\end{glossary}

\begin{abstract}[Abstract]
Quantum chromodynamics (QCD) with a general number of colors, $\Nc$, provides a powerful theoretical laboratory 
to explore the dynamics of non-Abelian gauge theories.
Although $\Nc =3$ does not look a large number, 
the $1/\Nc$ expansion provides us with a very useful classification and book-keeping scheme for hadronic processes 
and sharpens conceptions otherwise obscured in real-world QCD with $\Nc = 3$.
Important applications are dense QCD matter where the first principle methods for QCD are not available
and many conceptual issues remain to be clarified.
In this chapter we first review hadrons at large $\Nc$ from the viewpoint of quark-gluon dynamics,
and then extend the discussions to hot/dense matter, focusing on confinement-deconfinement aspects.
We emphasize how the large-$\Nc$ limit provides a unified organizing principle
for hadronic and quark degrees of freedom in regimes where first-principle methods are limited.
Two-color and isospin QCD, for which lattice simulations at finite density can be performed for a special reason,
is reviewed.
\end{abstract}

\section{Introduction}\label{sec:intro}

Quantum chromodynamics (QCD) with a general number of colors, $\Nc$, provides a powerful theoretical laboratory 
to explore the dynamics of non-Abelian gauge theories, 
including confinement, chiral symmetry breaking, and the structure of hadrons.
By varying the number of colors $\Nc$, one can uncover systematic patterns and scaling laws 
that are otherwise obscured in real-world QCD with $\Nc=3$.
This strategy was originally introduced by 't Hooft \citep{tHooft:1973alw}.

Although $\Nc =3$ does not look a large number, 
the $1/\Nc$ expansion provides us with a very useful classification scheme for hadronic processes 
and in practice it often works remarkably well.
This seemingly unreasonable success suggests that the coefficients of the $1/\Nc$ expansion
are numerically small in many physically relevant situations.
A similar situation occurs in quantum electrodynamics;
although the elementary charge $e \simeq 0.27$ is not very small,
in loop computations the coupling $e$ appears in powers of $\alpha_e = e^2/4\pi \simeq 1/137$, a very small number \citep{Witten:1979pi,Coleman:1985rnk}.
Another example is the critical exponent $\nu$ in the 3D Ising model;
in the expansion $\epsilon = 4-d$ ($d=3$: spatial dimension), one gets $\nu = 1/2 + \epsilon/12 + O(\epsilon^2)$
with which $\nu(\epsilon=1) \simeq 0.583$, reasonably close to the experimental value $\nu_{\rm exp} \simeq 0.630$ \citep{Wilson:1971dc}.
In this spirit, it would even make sense to study $\Nc=2$ theories as approximate laboratories for the real-world QCD.

An expansion in powers of $1/\Nc$ is a method of utility to classify various processes in hadron physics \citep{Callan:1975ps,Witten:1979pi}.
This can be done by matching the quark-gluon graphs with hadronic graphs.
The matching explicitly takes into account the fact that quarks are constituents of hadrons and also mediators of hadron-hadron interactions.
From this viewpoint, there should be a unified framework for the hadronic structure and hadron-hadron interactions.

In terms of $1/\Nc$ expansion, 
the decay width of a meson, $\Gamma$, divided by its mass $m$ so as to cancel the phase space factor,
is estimated to be $\Gamma/m \sim 1/\Nc$.
Thanks to this small width compared to its mass, one can treat mesons as if quasi-particles; if the width were as large as the mass, 
there would be no point to construct effective models of those very unstable mesons.
This sort of counting may be extended to the estimate for meson-meson interactions,
which are $\sim \Nc^{-1/2}$ for three-meson vertices, $ \sim \Nc^{-1}$ for four-meson vertices, and so on.
This leads to a picture that mesons are weakly interacting hadrons.

The application to the baryon sector is less straightforward than the mesonic sector,
but it still provides us with a useful guide \citep{Witten:1979pi}.
In particular, the strong channel dependence of baryon-baryon interactions 
can be vividly characterized in powers of $\Nc$ \citep{Kaplan:1995yg}.
Provided that the quark-meson coupling is $\sim \Nc^{-1/2}$,
the baryon-meson coupling depends on how the quark-meson couplings are assembled.
When $\Nc$-quarks contribute constructively, the baryon-meson coupling is $\sim \Nc^{-1/2} \times \Nc \sim \Nc^{1/2}$ which is strong.
Meanwhile when $\Nc$-quarks contribute destructively, the baryon-meson coupling is $\sim \Nc^{-1/2} \times 1 \sim \Nc^{-1/2}$.
In terms of baryon-baryon forces, the overall strength is proportional to the square of the vertices,
so the former is $\sim \Nc$ and the latter is $\sim 1/\Nc$; there is a factor $\Nc^2\sim 10$ difference.
This large hierarchy is indeed seen in the meson exchange potentials (e.g., \cite{Stoks:1994wp}).

The $1/\Nc$ expansion provides particularly valuable guidance in discussing QCD matter at finite temperature and density.
One of contemporary problems in quantum chromodynamics (QCD) is to understand transitions from a matter of hadrons 
to those of quarks and gluons (for a review, e.g., \cite{Fukushima:2010bq}).
These hadronic estimates are building blocks to consider such transitions.
In the dilute regime, quarks and gluons are confined into hadrons so that the properties of a matter may be understood in terms of the hadronic language.
This situation changes in the hot/dense regime where 
hadrons strongly interact and/or spatially overlap so that quarks and gluons become natural effective degrees of freedom to characterize the medium.
Many key questions in hot and dense QCD are associated with this intermediate regime,
where neither purely hadronic nor weakly coupled quark-gluon descriptions are sufficient.

Expanding our scope to QCD with a general number of colors, 
two-color QCD (QC$_2$D) provides a theoretical laboratory
for studying nonperturbative aspects of QCD.
One of the distinct properties of this system is that, unlike real-world QCD,
lattice Monte Carlo simulations can be performed at finite baryon density \citep{Hands:1999md}.
In fact, over the last decade, lattice simulations of dense QC$_2$D
have confirmed several concepts of dense matter
that were originally inferred from neutron star observations \citep{Iida:2022hyy}.
Although there is an important difference between QC$_2$D and QCD in that
a baryon is a boson in the former and a fermion in the latter,
in sufficiently dense regimes where baryons overlap and quark degrees of freedom become manifest,
it is reasonable to expect that the resulting matter shares qualitative similarities in these two theories.
A similar strategy can also be applied to QCD with a finite isospin chemical potential (QCD$_I$),
for which lattice simulations are likewise applicable \citep{Brandt:2022hwy,Abbott:2023coj}.

In this chapter, we will review the basics and applications of QCD with a general $\Nc$.
After briefly summarizing the basics of $1/\Nc$ expansion,
we apply the framework to classifications of hadronic processes.
The hadronic estimates are subsequently used to study many-hadron systems,
and then extended to the studies of hot/dense matter.
We also discuss QC$_2$D and QCD$_I$ by referring to recent lattice results and model studies.
Throughout this chapter, we emphasize qualitative patterns and physical intuition, and refer the reader to the original literature for technical details.

\section{The $1/\Nc$ expansion: definition } \label{sec:nc}

\begin{figure}[t]
\vspace{-1.5cm}
\centering
\includegraphics[width=.6\textwidth]{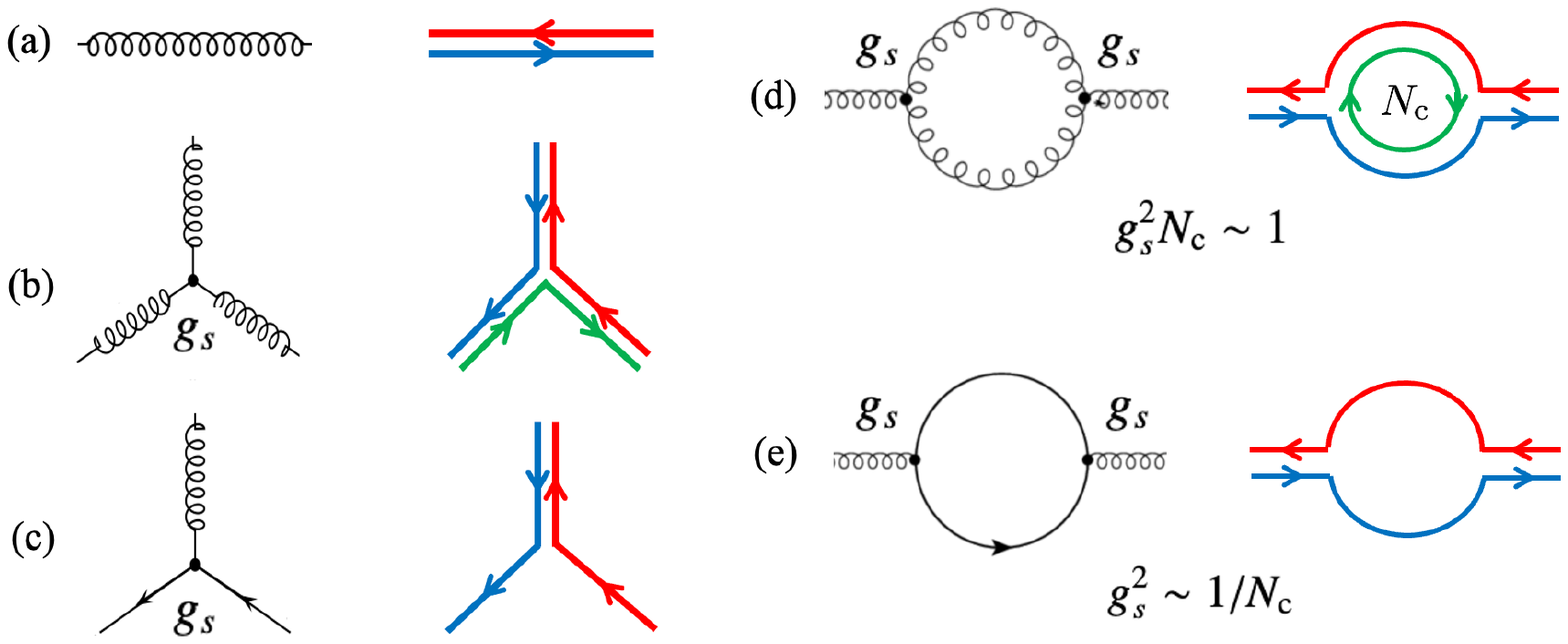}
\vspace{-1.2cm}
\caption{
Double-line representation and large-$\Nc$ counting rules.
(a)--(c) Fundamental interaction vertices.
(d) Planar gluon loop giving a factor $\Nc$ with $g_s^2 \Nc \sim 1$ fixed.
(e) Non-planar contraction suppressed as $g_s^2 \sim 1/\Nc$.
}
\label{fig:Nc_graphs}
\end{figure}

In 't Hooft large $\Nc$ limit, we take $\Nc$ large while holding $\lambda = \Nc g_s^2$ fixed \citep{tHooft:1973alw}.
Here $g_s$ is the gauge coupling constant.
This condition on $g_s^2$ comes from the requirement
that the running $\alpha_s = g_s^2/4\pi$ in large $\Nc$ QCD behaves in the similar way as QCD.
Explicitly, the running $\alpha_s$ at one-loop level scales as
\beq
\alpha_s (Q^2) = \frac{\, 4\pi \,}{\, \beta_0 \ln \big(Q^2/\lqcd^2 \big) \,}\,,~~~~~~~
\beta_0 = \frac{\, 11\Nc - 2 \Nf \,}{3} \,,
\eeq
where $\Nf$ is the number of flavors and $\lqcd \simeq 200$-300 MeV is the typical nonperturbative scale in QCD.
In QCD, there are $\sim \Nc^2$ gluons and $\sim \Nc \Nf$ quarks.
In the large $\Nc$ limit $(\Nc \gg \Nf)$, the dynamics is dominated by gluons
and the impact of quarks are treated as corrections.

It is convenient to use color line representation for the Feynman diagrams.
The gluon fields $A_a$ $(a=1,\cdots \Nc^2-1)$ as the adjoint representations of $SU(\Nc)$ may be rewritten as
$A^{i}_j \equiv A_a (T_a)^{i}_j$ where matrices $T_a$ form $SU(\Nc)$ algebra 
and one may use the combinations of $i, j =1,\cdots, \Nc$ to specify the gluon species.
Using the indices $i,j$, one can keep track of the color charges which are conserved.
For instance, a gluon propagator can be written as (Fig.~\ref{fig:Nc_graphs} (a)).
\beq
\la (A_\mu)^{i}_j (x) (A_\nu)^k_l (0) \ra
= \la A_\mu^a (x) A_\nu^b (0) \ra \times (T_a)^{i}_j (T_b)^k_l
= D_{\mu \nu} (x) \times (T_a)^{i}_j (T_a)^k_l
=  D_{\mu \nu} (x) \times \frac{1}{\, 2 \,} 
\bigg( \delta^{i}_l \delta_j^k - \frac{1}{\, \Nc \,}  \delta^{i}_j \delta_l^k \bigg) \,,
\eeq
where we used $ \la A_a (x) A_b (0) \ra = D(x) \delta_{ab}$.
The Kronecker $\delta$ in the last expression reflects how colors are transferred.
The last $1/\Nc$ term comes from the difference between $SU(\Nc)$ and $U(\Nc)$.
This difference is usually negligible.

Likewise, the color conservation at three-gluon vertices can be represented as
(Fig.~\ref{fig:Nc_graphs}(b))
\beq
f_{abc} \big( \partial_\mu A_\nu^a \big) A_\alpha^b A_\beta^c 
= 2 \rmi (T_a)^i_j \bigg[ (T_b)^j_k (T_c)^k_i - (T_c)^j_k (T_b)^k_i \bigg] 
	\times \big( \partial_\mu A_\nu^a \big) A_\alpha^b A_\beta^c  
= 2 \rmi 
\big( \partial_\mu A_\nu \big)^i_j \bigg[ 
	\big( A_\alpha \big)^j_k \big( A_\beta \big)^k_i  	
	- \big( A_\beta \big)^j_k \big( A_\alpha \big)^k_i  	
\bigg] \,.	
\eeq
The four-gluon vertex can be discussed in the similar way.
As for the color transfer between gluons and quarks,
the usual quark-gluon coupling may be rewritten as
(suppressing the $\gamma$-matrices, Fig.~\ref{fig:Nc_graphs}(c))
\beq
\bar{q} A_a T_a q
= \bar{q}_i \partial A_a \big( T_a \big)^i_j q^j
= \bar{q}_i A^i_j q^j \,.
\eeq
From the contraction of the indices,
one can see how the colors carried by quarks and antiquarks are transferred to gluons.
After establishing the color conservation, 
one can assign $\Nc$ counting for various graphs for interaction processes.

To discuss loop corrections, we first consider graphs including only gluon lines (e.g., Fig.~\ref{fig:Nc_graphs}(d) for one-gluon loop).
Since $g_s \sim \Nc^{-1/2}$, increasing the order of $g_s$ introduces more suppression factors.
Meanwhile, in loop computations there appear graphs
in which color lines are contracted by themselves, forming loops.
Since loops may be formed for any colors,
the color loops introduce an enhancement factor $\Nc$.
When we consider the so-called planar graphs in which all color lines may be drawn on a plane,
at any order of loops $g_s^2 $ appears together with the enhancement factor $\Nc$,
so that $g_s^2 \Nc \sim 1$ contribute at the same order of $\Nc$ as graphs with less number of loops.
That is, all planar graphs contribute equally in terms of the $1/\Nc$ expansion.
Meanwhile, if the graphs contain non-planar lines,
we have less color line contraction and hence lose the enhancement factor $\Nc$, see \cite{tHooft:1973alw} for those examples.
Those graphs are regarded as sub-leading.

Next we include quarks  (Fig.~\ref{fig:Nc_graphs}(e)).
It is instructive to consider one loop correction from gluons and quarks to a gluon propagator.
As we have just discussed above, gluon loops connect one index with the external line
and contract the other index to generate a color loop with the $\Nc$ enhancement.
A quark loop does not have such $\Nc$ enhancement,
and hence is smaller than the gluon loop by a factor $\sim 1/\Nc$.
Every time we replace a gluon loop with a quark one,
the graph acquires the suppression factor $\sim 1/\Nc$.

\section{ Mesons at large $\Nc$ } \label{sec:mesons}

\begin{figure}[t]
\vspace{-1.5cm}
\centering
\includegraphics[width=.6\textwidth]{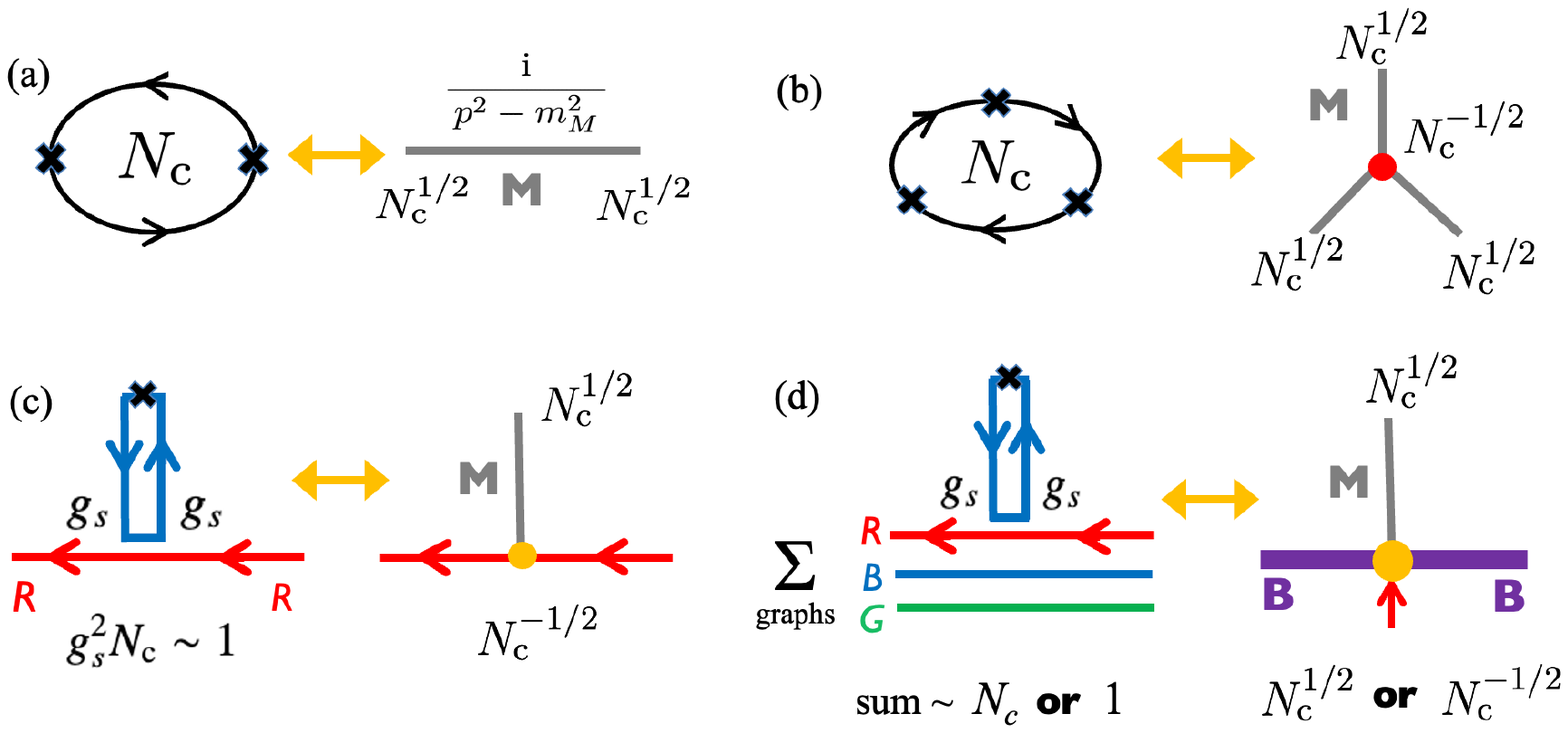}
\vspace{-1.cm}
\caption{ 
Large $\Nc$ matching between quark-gluon and hadronic graphs.
(a) Matching of the meson two-point function, which fixes the
state-operator coupling
$\la M | \bar{q}\Gamma q | 0 \ra \sim \Nc^{1/2}$.
(b) Meson three-point interaction whose vertex scales as $\Nc^{-1/2}$.
(c) Quark-meson coupling determined to be $\Nc^{-1/2}$.
(d) Baryon-meson coupling scaling as $\Nc^{1/2}$ or $\Nc^{-1/2}$,
depending on whether quark--meson graphs add constructively or destructively.
}
\label{fig:meson_baryon_Nc}
\end{figure}

Based on the $\Nc$ counting in the previous section,
we now consider matching between quark-gluon graphs and hadronic graphs.
Our first step is to consider a correlation function 
$\la 0 | J_M (x) J_M^\dag (0) | 0 \ra$ with $J_M$ being a color singlet quark bilinear operator $\bar{q} \Gamma q $ (Fig.~\ref{fig:meson_baryon_Nc}(a)).
In the color line representation,
the correlation function includes a color loop and is estimated to be $\sim \Nc$.
On the other hand, in the hadronic language $J_M^\dag$ creates a hadronic state $|n\ra$,
and this state evolves in time, and then get annihilated by the operator $J_M$.
Taking the Fourier transform, our matching can be expressed as
\beq
\int_x \rme^{\rmi p x} \la 0 | J_M (x) J_M^\dag (0) | 0 \ra 
~\sim~ \Nc
~\sim~ \sum_n \la 0 | J_M (0) | n (\vp) \ra \frac{\, \rmi \,}{\, p^2 - m_n^2 \,} \la n  (\vp) | J_M^\dag (0) | 0 \ra \,.
\eeq
From this matching, one can estimate the state-operator coupling $\la n  (\vp) | J_M^\dag (0) | 0 \ra$ to be $\sim \Nc^{1/2}$.

Our next step is to examine the $\Nc$ counting for meson vertices.
We begin with three meson vertices  (Fig.~\ref{fig:meson_baryon_Nc}(b)).
We consider a three point function among three interpolating fields $J_{M_i}$ with $i=1,2,3$.
In quark-gluon graphs, this simply forms a single color loop and hence is estimated to be $\sim \Nc$.
To match this counting in terms of hadronic graphs,
each state-operator coupling is $\sim \Nc^{1/2}$ so that
the three meson vertex must scale as $\Nc^{-1/2}$.
With this weak coupling, processes such as a decay process $M_1 \rightarrow M_2 M_3$ or a fusion process $M_1 M_2 \rightarrow M_3$
are suppressed at large $\Nc$.

More quantitatively,
squaring the former amplitude for a meson $n$, we estimate the width $\Gamma_n$ to be $\sim 1/\Nc$ times a factor depending on the phase space available for decays.
Since the phase space factor depends on the overall mass scale,
it is appropriate to look at the ratio $\Gamma_n/m_n$ where the phase space factor is largely cancelled.
For instance, the ratio for $\rho(770)$ with $m_\rho \simeq 775$ MeV and $\Gamma_\rho \simeq 150$ MeV is $\Gamma_\rho/m_\rho \sim 0.2$;
for $\omega (782)$, $\Gamma_{\omega}/m_{\omega} \sim 0.011$; for $K^*(892)$, $\Gamma_{K^*}/m_{K^*} \sim 0.06$.
Even $a_1(1260)$ with the apparently large decay width $\Gamma_{a_1} \sim 250$-$600$ MeV,
the ratio $\Gamma_{a_1}/m_{a_1} \sim 0.2$-$0.5$ is not so bad.
These small ratios $\Gamma_n/m_n \sim 1/\Nc \ll 1$ 
allow us to regard mesons as quasi-particles
and to employ an effective mesonic Lagrangian as an approximate framework.
(Actually there are also exceptional mesons which significantly violate these estimates ---
they are often called exotics, see, e.g., \cite{Pelaez:2003dy,Weinberg:2013cfa}).

Next we consider four meson vertices. The straightforward extension of the three vertex case
allows us to estimate the four-meson vertex to be $\Nc^{-1}$.
Every time we attach an additional meson line, the strength of the vertex is suppressed by an extra factor $\Nc^{-1/2}$.
As a result, meson theories in the large $\Nc$ limit become weakly interacting theories.
%

\section{ Baryons at large $\Nc$ } \label{sec:baryons}

The utility of the $1/\Nc$ expansion in the baryon sector
is less firmly established than in the meson sector.
In particular, baryons have the masses of $\sim \Nc$,
and their internal structure and interactions are more sensitive
to dynamical details.

\subsection{ Baryon-meson coupling } \label{sec:BMcoupling}

In this section, we begin with baryon--baryon interactions.
The classification of baryonic interactions in powers of $1/\Nc$
provides a useful organizing principle \citep{Witten:1979pi}
and captures several qualitative aspects of nuclear forces \citep{Kaplan:1995yg}.

First we note that the quark-meson coupling is $\Nc^{-1/2}$ (Fig.~\ref{fig:meson_baryon_Nc} (c)).
To see this, we consider a diagram in which an interpolating field $J_M$
couples to two external quark lines.
For simplicity we consider a graph with one gluon exchange.
There is a color loop yielding a factor $\Nc$,
and two quark-gluon vertices introduce $g_s^2 \sim \Nc^{-1}$ so that
this graph is $\sim 1$ as a total.
As a hadronic graph, the state-operator coupling is $\sim \Nc^{1/2}$
so that the quark meson coupling must be $\sim \Nc^{-1/2}$.

To estimate baryon-meson couplings, 
we sum up graphs in which a meson line is attached to one of $\Nc$ quarks in a baryon (Fig.~\ref{fig:meson_baryon_Nc} (d)).
The consequence of the summation strongly depends on the channel we consider.
The baryon-meson coupling is strong
when each quark contribution is added constructively;
in this case there is a factor $\Nc$ enhancement so that the baryon-meson coupling constant is large,
$g_{BBM}^{\rm constructive} \sim \Nc^{-1/2} \times \Nc \sim \Nc^{1/2}$.
Meanwhile, there are also channels in which
quark contributions are added destructively;
in this case there is no $\Nc$ enhancement and the baryon-meson coupling constant remains small,
$g_{BBM}^{\rm destructive} \sim \Nc^{-1/2} \times 1 \sim \Nc^{-1/2}$.
In meson exchange potentials,
its strength is $\sim g_{BMM}^2$,
so that the ratio $1/\Nc$ in couplings is enlarged to the $1/\Nc^2$ difference in the corresponding potentials.

A useful example is the difference between $\omega$ and $\rho$ exchange potential between two nucleons.
They appear in the form $\sim g_{\omega NN} \omega_0 \bar{N} \gamma_0 N$ and $\sim g_{\rho NN} \rho_0^a \bar{N} \gamma_0 \tau_a N$.
Here we consider the temporal or electric components of these mesons for which we use 
$\omega_0$ and $\rho_0$ in our notation.
The $\omega_0$ couples to the quark numbers  carried by the baryon
and there is $\Nc$ enhancement since
all quarks have positive quark/baryon number.
The situation differs for the $\rho_0$ meson,
as it couples to the isospin charge of a nucleon which is $\sim 1$ in the $\Nc$ counting.
Nucleons have both $u$- and $d$-quarks and they have opposite isospins,
so that contributions with the opposite signs cancel.
More quantitatively, in a representative meson exchange potential (e.g., \cite{Stoks:1994wp}),
the ratio $g_{\omega NN}/g_{\rho NN}$, evaluated so as to cancel the form-factor dependence, is found to be
$g_{\omega NN}/g_{\rho NN}  \sim 3.3$.
This value is consistent with the expectation from the large $\Nc$ counting.

Another illustrative example is given by the $\sigma$ and $a_0(980)$ exchange
contributions to the nucleon–nucleon interaction.\footnote{
Here $\sigma$
denotes an effective scalar–isoscalar degree of freedom commonly employed
in meson-exchange descriptions of nuclear forces, rather than a narrow fundamental resonance.
} 
The corresponding interaction terms take the form
$\sim g_{\sigma NN}\,\sigma\,\bar{N}N$ and
$\sim g_{a_0 NN}\,a_0^a\,\bar{N}\tau_a N$.
These mesons couple to the Lorentz scalar density in the isoscalar and isovector channels,
respectively.
According to large-$\Nc$ counting, the couplings scale as
$g_{\sigma NN}\sim \Nc$ and $g_{a_0 NN}\sim \mathcal{O}(1)$.
More quantitatively, meson–exchange potentials \citep{Stoks:1994wp} yield
$g_{\sigma NN}/g_{a_0 NN} \simeq 3.8$,
which is consistent with this large-$\Nc$ expectation.

%

\subsection{ The nucleon axial charge } \label{sec:axial_charge}

\begin{figure}[t]
\vspace{-2.5cm}
\centering
\includegraphics[width=.7\textwidth]{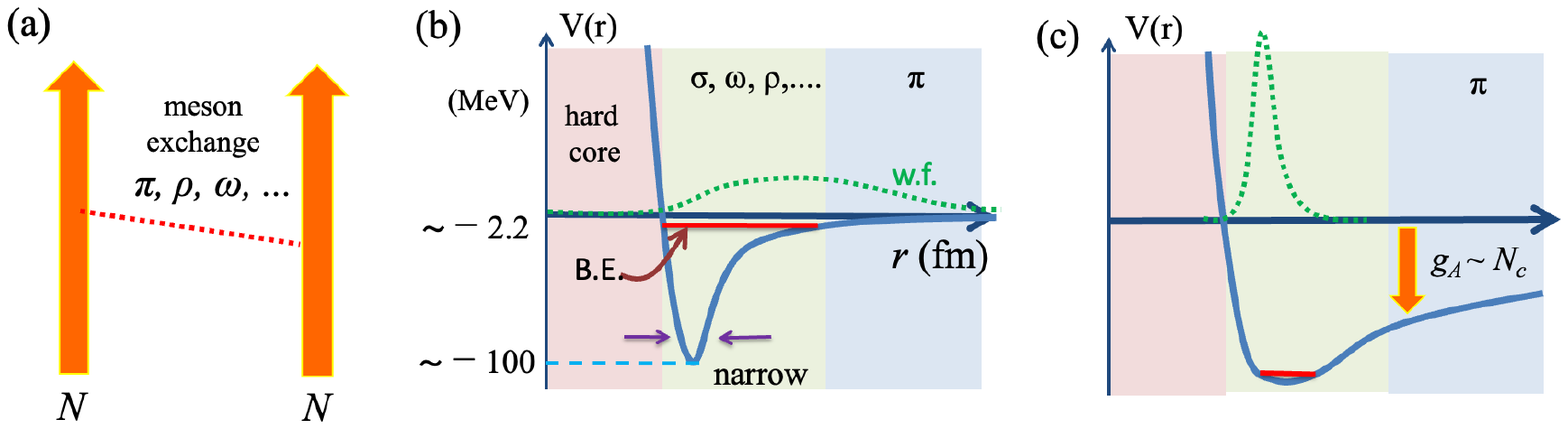}
\vspace{-2.2cm}
\caption{ 
Schematic nuclear forces and their large $\Nc$ implications.
(a) Meson-exchange picture of the nucleon-nucleon interaction.
(b) Qualitative shape of the nuclear potential at $\Nc=3$, showing
a hard core, intermediate-range attraction, and long-range pion exchange,
leading to a shallow bound state.
(c) Extrapolation to large $\Nc$, where the axial coupling
$g_A \sim \Nc$ enhances the long-range pion-exchange interaction,
leading to a deeply bound two-nucleon state.
As a consequence, nuclear matter is expected to form a crystal at large $\Nc$.
}
\label{fig:Nc_nuclear}
\end{figure}

The above discussions with vector and scalar mesons exchanges are relatively simple
as one can estimate the (vector) charges and Lorentz scalar charges rather easily.
A more subtle and intriguing channel is the axial (-vector) charges to which pions couple.
In non-relativistic operator languages, the nucleon axial charge is proportional to the spin-isospin operator 
\beq
G_{ia} \equiv \sigma_i \tau_a \,,
\eeq
which, together with the spin and isospin generators, forms an $SU(4)$ spin-flavor algebra \citep{Gervais:1983wq,Dashen:1993jt}.
In the standard quark model, it is known that the ground-state baryons are described by a completely symmetric spin-flavor wave function
for which one finds $\la N | G_{ia} | N \ra \sim \Nc$ \citep{Manohar:1984ys}.
As a consequence, the axial coupling scales as 
$g_A \sim \Nc$, and the pion-nucleon coupling is parametrically enhanced.
The Goldberger-Treiman relation $g_{\pi NN} f_\pi = g_A m_N$ ($f_\pi$: pion decay constant, $m_N$: nucleon mass)
suggests $g_{\pi NN} \sim \Nc^{3/2}$ since $f_\pi \sim \Nc^{1/2}$ and $M_N \sim \Nc$ \citep{Hidaka:2010ph}.

Regarding $g_{\pi NN}$ as strong coupling leads to a picture of large pion clouds (coherent pions) surrounding the nucleon core.
Those pions take a specific field configuration, chiral soliton, with a topological charge one which is identified as a baryon number.
The most famous construction is the Skyrmion \citep{Skyrme:1961vq,Skyrme:1962vh} which is the classical solution of the Lagrangian
(for a review, \cite{Zahed:1986qz})
\beq
\calL 
= \frac{\, f_\pi^2 \,}{4} \tr\big( \partial_\mu U \partial^\mu U^\dag \big)
+ \frac{\, 1 \,}{\, 32 e^2 \,} \tr\bigg( \big[ U^\dag \partial_\mu U,  U^\dag \partial_\nu U \big]^2 \bigg)
+ \frac{\, f_\pi^2 m_\pi^2 \,}{4} \tr\big( U + U^\dag - 2 \big) \,,
~~~~~~~~~ U = \rme^{\rmi \pi_a \tau_a/f_\pi} \,.
\eeq
Assuming fields $\pi$ to be classical or large amplitude fields,
the quadratic and quartic derivative terms are balanced to yield soliton solutions.
The spin and isospin quantum number of this soliton is the mixture of several baryon species.
This is energetically possible since,
in conventional quark models at large $\Nc$, the nucleon and $\Delta$ mass splitting comes from the color magnetic interactions of $\alpha_s \sim 1/\Nc$,
meaning that their masses become degenerate at large $\Nc$.
Hence, in order to describe a nucleon, 
one needs to project the nucleon quantum number from the classical soliton (projection after variation) \citep{Adkins:1983ya}.
There are several upgraded versions of the Skyrme model which include vector mesons such as $\rho$ and $\omega$ mesons.
A more elegant construction is the instanton construction in holographic QCD with infinite towers of mesons 
included (e.g., \cite{Sakai:2004cn,Hata:2007mb,Hong:2007kx}).

\subsection{ Nuclear binding } \label{sec:nuclear_binding}

While the large $\Nc$ baryons and the resulting soliton construction include a lot of interesting physics,
the direct application to the nuclear physics is subtle.
In nuclear physics, there emerges an unnaturally small scale of a few MeV
which is typical in nuclear binding energies (Fig.~\ref{fig:Nc_nuclear}).
In nuclear matter near the saturation density, 
the kinetic energy of nucleon, $\sim \vp^2/m_N \sim \lqcd/\Nc$, is typically several tens of MeV,
while the binding energy seems to be $\sim \lqcd/\Nc^2$.
For this to happen, the kinetic and potential energies should largely cancel,
so the potential energy should be $\sim \lqcd/\Nc$, especially at long distance.
If we consider the large pion fields of $\sim \Nc$, 
the attractive potential at long distance is too large.
Pushing this description to many-body systems leads to crystals of chiral solitons \citep{Klebanov:1985qi}, 
since the potential dominates over the kinetic energy and 
it is energetically more favorable for nucleons to take the fixed locations.
In contrast, the empirical nuclear matter is like a liquid
where the kinetic and potential energies are comparable.

To summarize this section,
the $1/\Nc$ classifications for baryon-baryon interactions work in many channels,
but the channel involving the axial charges needs special care.
This issue propagates to descriptions of many-body problems.
In practice, treating $g_A$ as order unity provides a reasonable picture for dense nuclear matter \citep{Hidaka:2010ph}.
In the following, we proceed to denser matter within the picture of $g_A\sim 1$, without rigorous justifications.

\section{ Hot matter at large $\Nc$ } \label{sec:hot_matter}

Now we extend the $1/\Nc$ expansion to hot matter.
The state of matter can be specified by thermodynamic variables, baryon chemical potential $\mu_B$ and temperature $T$,
and the pressure $P (\mu_B, T) = \mu_B n_B + T s - \varepsilon $
($n_B$ baryon density; $s$: entropy density; $\varepsilon$: energy density).
Our understanding of the hot regime of QCD has matured,
thanks to the combined use of lattice QCD, heavy-ion experiments, and physics-inspired 
effective models (for reviews, e.g., \cite{BraunMunzinger:2016,Shuryak:2014zxa}).
Equations of state, transport properties, hadronic correlations in hot medium, and so on,
have been examined in detail.

First we consider a hot matter at $\mu_B = 0$.
At finite temperature, 
the abundance of thermally excited hadrons 
is controlled by the Boltzmann factor $\rme^{-\beta E_h}$ ($\beta=1/T$: inverse temperature)
times the degeneracy factor including spins, isospins, and so on.
At large $\Nc$, masses of mesons and glueballs are $\sim \Nc^0 \lqcd$, while baryon masses are $\sim \Nc \lqcd$.
Thus, at $T \ll \Nc \lqcd$, a hot matter is dominated by a gas of mesons and glueballs in which interactions are suppressed by $1/\Nc$.
Since the decay width of mesons and glueballs are suppressed by $1/\Nc$,
these particles may be regarded as quasi-particles. 
This allows us to write the pressure simply as
\beq
P (T) = - T \sum_n \int_{\vp} \ln \big( 1 - \rme^{-\beta E_n (\vp) } \big) 
\,,~~~~~~ E_n (\vp) = \sqrt{ \vp^2 + m_n^2 } \,,
\eeq
where $n$ characterizes a hadronic state with the mass $m_n$.
We sum over all possible hadronic states as independent.
There are no quantities which manifestly depend on $\Nc$, and hence the resultant pressure is $\sim \Nc^0$.

At large excitation energies, the number of hadronic states grows rapidly because there are many possible stringy excitations,
i.e., excitations of color-electric flux connecting colored objects such as quarks and gluons.
There are open strings connecting quarks and antiquarks, and also closed strings making glueball states.
At large $\Nc$, string breaking and string interactions are suppressed,
so that these string excitations can be counted as independent states.
The string models for hadrons suggest that the density of hadronic states, or entropy, exhibits an exponential growth,
\beq
\rho(m) \sim m^{-a} \rme^{m/T_H} \,, 
\eeq
where $T_H$ is called the Hagedorn temperature \citep{Hagedorn:1965st}. Replacing the discrete sum over hadronic states by an integral over the mass,
the pressure takes the schematic form
\beq
P(T) \sim - \int \rmd m \, \rho(m) \int_{\vp}
\ln \big( 1 - \rme^{-\beta \sqrt{\vp^2+m^2}} \big) 
\sim 
\int \rmd m ~ m^{-a} \rme^{ m \big( \frac{1}{\, T_H \,} - \frac{1}{\, T \,} \big) }
\int_{\vp} \rme^{- \beta \frac{\vp^2}{\, 2m \,}+ \cdots }
\,.
\eeq
As $T$ approaches $T_H$ from below, the exponential growth of $\rho(m)$
overcomes the Boltzmann suppression, and the partition function ceases to be well-defined.

This ill-defined behavior comes from overcounting of states;
we have treated all hadrons and the associated strings as independent.
When hadrons or the associated color flux strongly overlap, however,
we should imagine the condensation of the color fluxes that overcomes the suppression factor $1/\Nc$ \citep{Polyakov:1978vu,Gross:1980br}.
Such a matter forms a colored background within which quarks and gluons directly contribute to the thermodynamics,
see quark models augmented by Polyakov loops as successful examples \citep{Fukushima:2003fw,Ratti:2005jh,Schaefer:2007pw,Fukushima:2017csk}.
In this sense, the Hagedorn temperature may be viewed as a limiting temperature of the hadronic description.
Once the colored background is formed, the pressure is dominated by gluons of $\sim \Nc^2$ which is regarded as bigger than quark contributions of $\sim \Nf \Nc$.
This pressure is substantially larger than the hadronic pressure of $\sim \Nc^0$,
discriminating the confined and deconfined phases.
If quark contributions are neglected, one finds the deconfinement temperature $T_{\rm deco} \sim 270$ MeV.

While the above large $\Nc$ descriptions capture the qualitative trends of confinement-deconfinment transitions,
there are substantial corrections at $\Nc = 3$.
The quantitative aspects have been examined by lattice Monte-Carlo results.
Indeed, it is established that the transition from hadronic to QGP phase is smooth crossover \citep{Aoki:2006we}
which begins at the pseudo-critical temperature $T_c \simeq 155$ MeV \citep{Bhattacharya:2014ara}.
Around $T_c$, the entropy density is $s \simeq 2$-3 fm$^{-3}$, 
suggesting that two- or three-quark composites, mesons and baryons with the radii $\sim 0.5$-1 fm, begin to overlap.
This is supported by the fact that
the hadron resonance gas model, including substantial baryon contributions at $\Nc=3$,
works very well to $T_c$ \citep{Karsch:2001nf,Huovinen:2009yb}, 
but beyond which the pressure (entropy) is overestimated.
Various number susceptibilities including those of charm quarks begin to change at the common temperature $T_c$ \citep{Ding:2016cqn};
this trend is consistent with the picture of the string condensation at large $\Nc$ which form a background universal for all flavors. 
What is specific at $\Nc=3$ is that the color-flux is mainly brought by mesons and baryons.

To summarize, while the large-$\Nc$ descriptions are not quantitatively precise,
they provide very useful qualitative guidelines and organizing principles for hot QCD.

\section{ Cold, dense matter at large $\Nc$ } \label{sec:dense}

In the cold, dense regime, 
a matter is formed when the baryon chemical potential
$\mu_B = \Nc \mu_q$ ($\mu_q$: quark chemical potential) 
exceeds the nucleon mass $m_N \sim \Nc \lqcd$ (minus the binding energy),
or equivalently when $\mu_q$ exceeds
the effective quark mass $M_q \sim \lqcd$.
As announced in Sec.~\ref{sec:nuclear_binding}, 
we assume that nuclear matter remains liquid even at large $\Nc$, provided that $g_A \sim 1$.

We first quickly review findings since 2010 for cold, dense matter in the real-world QCD.
The major sources of empirical information 
are the nuclear physics around saturation density $n_0 \simeq 0.16\,{\rm fm}^{-3}$
and neutron star constraints.
In particular, the mass-radius $M$-$R$ measurements for neutron stars, which have drastically improved since the discovery of a two-solar mass neutron star, 
have provided a lot of information for the QCD equations of state (EOS) for $n_B = 1$-$5n_0$.
We then discuss several concepts deduced from large $\Nc$ arguments
which are useful to formulate theoretical problems.

\subsection{ Implications from empirical constraints } \label{sec:EOS}

\begin{figure}[t]
\vspace{-1.5cm}
\centering
\includegraphics[width=.6\textwidth]{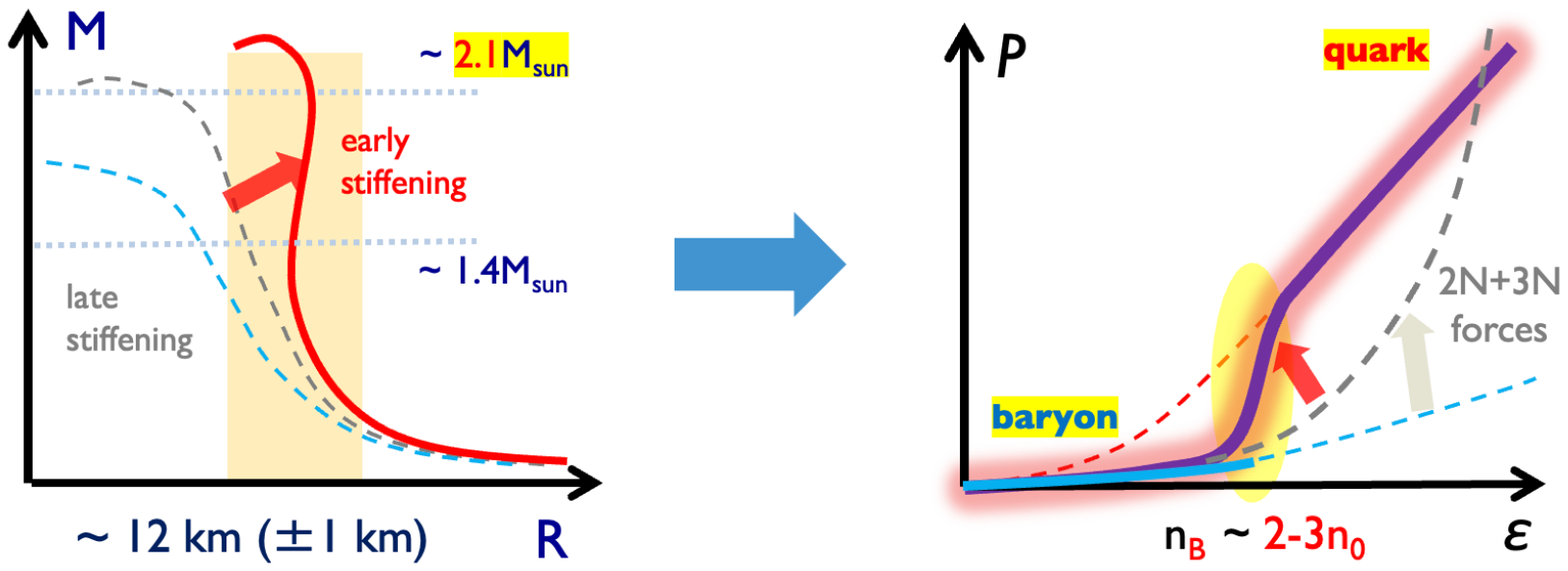}
\vspace{-1.5cm}
\caption{ 
Observational constraints on neutron star $M$-$R$ relations and the corresponding inference which suggests that
the EOS stiffens rapidly around 2-3$n_0$ ($n_0\simeq 0.16\,{\rm fm}^{-3}$: nuclear saturation density $\simeq$ nucleon density in typical nuclei) 
and approaches the quark matter behavior with $P \simeq \varepsilon/3$.
}
\label{fig:EOS_MR}
\end{figure}

EOS of matter and $M$-$R$ relations of neutron stars have one-to-one correspondence.
If the $M$-$R$ curve is determined precisely, one can directly obtain the EOS
and from which we can infer the microphysics of dense matter \citep{Lattimer:2000nx,Alford:2006vz,Bauswein:2018bma,Baym:2017whm,Annala:2019puf}.

One of the most important concepts in neutron star EOS is the stiffness of matter
or the relative magnitude of pressure ($P$) with respect to the energy density ($\varepsilon$).
The energy density is the source of gravitational contraction
while the pressure counteracts on the compression of matter.
A stiff (soft) EOS has a large (small) $P$ at a given $\varepsilon$;
when $P$ grows more rapidly (slowly) with increasing $n_B$ and $\varepsilon$,
it means that the matter is harder (easier) to be compressed. 
For instance, a matter of non-relativistic baryons leads to $\varepsilon \sim m_N n_B + n_B^{5/3}/m_N$,
and the resulting pressure is $P = n_B^2 \partial (\varepsilon/n_B)/\partial n_B \sim n_B^{5/3}/m_N$, suppressed by its large mass.
Hence $P \ll \varepsilon$, very soft.
For a matter of relativistic particles, $\varepsilon \sim n_B^{4/3}$, the pressure is $P \sim n_B^{4/3} \sim \varepsilon$ and hence the EOS is stiff.
In the relativistic limit, one can find $P =\varepsilon/3$ after eliminating $n_B$ in favor of $\varepsilon$.

One of symbolic findings in recent neutron star observations is the peak in the (adiabatic) sound speed, 
$c_s = \big( \partial P/\partial \varepsilon \big)^{1/2}$,
exceeding the value of the relativistic limit, $c_s = (1/3)^{1/2} $ \citep{Masuda:2012ed,Kojo:2014rca,Bedaque:2014sqa,Ma:2019ery}.
The Bayesian inference, constrained by nuclear physics, neutron star observations, and perturbative QCD at very high density,
suggests that the peak appears at densities intermediate between baryonic and quark matter, 
$n_B \sim 2$-$5n_0$ (e.g., \cite{Brandes:2022nxa,Han:2022rug}). 
If we consider transitions from baryonic to quark matter as distinct {\it phase} transitions,
it is difficult to explain the sound speed peak or the rapid stiffening.
Indeed, at first order phase transitions, the pressure is constant but the energy density jumps discontinuously,
leading to $c_s = 0$.
For this reason, the possibility of the quark-hadron crossover has been explored as a viable scenario \citep{Masuda:2012ed,Baym:2017whm}.
Explaining rapid stiffening from the baryonic to quark matter regimes is one of key issues in the contemporary dense QCD physics.

\subsection{ Quarkyonic matter } \label{sec:quarkyonic}

\begin{figure}[t]
\vspace{-2.5cm}
\centering
\includegraphics[width=.6\textwidth]{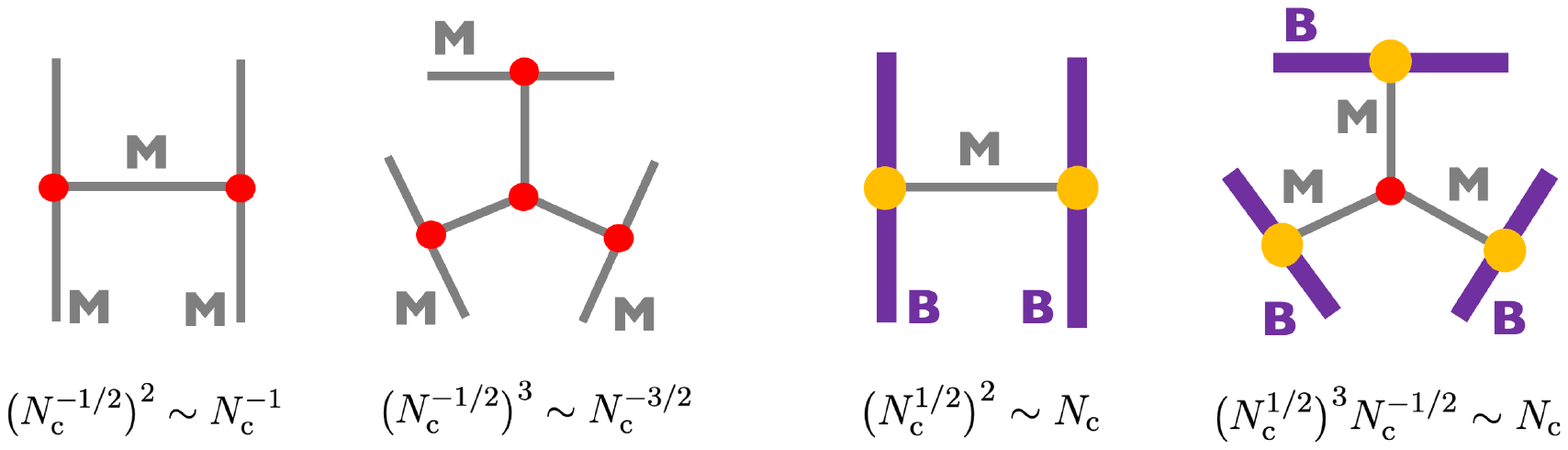}
\vspace{-2.cm}
\caption{ 
Large-$\Nc$ scaling of many-body forces in mesonic and baryonic systems.
For mesons, the $N$-body vertices scale as $\sim \Nc^{-N/2}$ and are increasingly
suppressed for larger $N$, implying that hot mesonic matter is dominated by
two-body interactions.
For baryons, in contrast, the $N$-body vertices can remain as large as the
two-body vertices, scaling as $\sim \Nc$.
As a result, many-body forces play a crucial role in dense baryonic matter,
in sharp contrast to mesonic matter.
%
}
\label{fig:many-M_B}
\end{figure}

\begin{figure}[t]
\vspace{-1.cm}
\centering
\includegraphics[width=.6\textwidth]{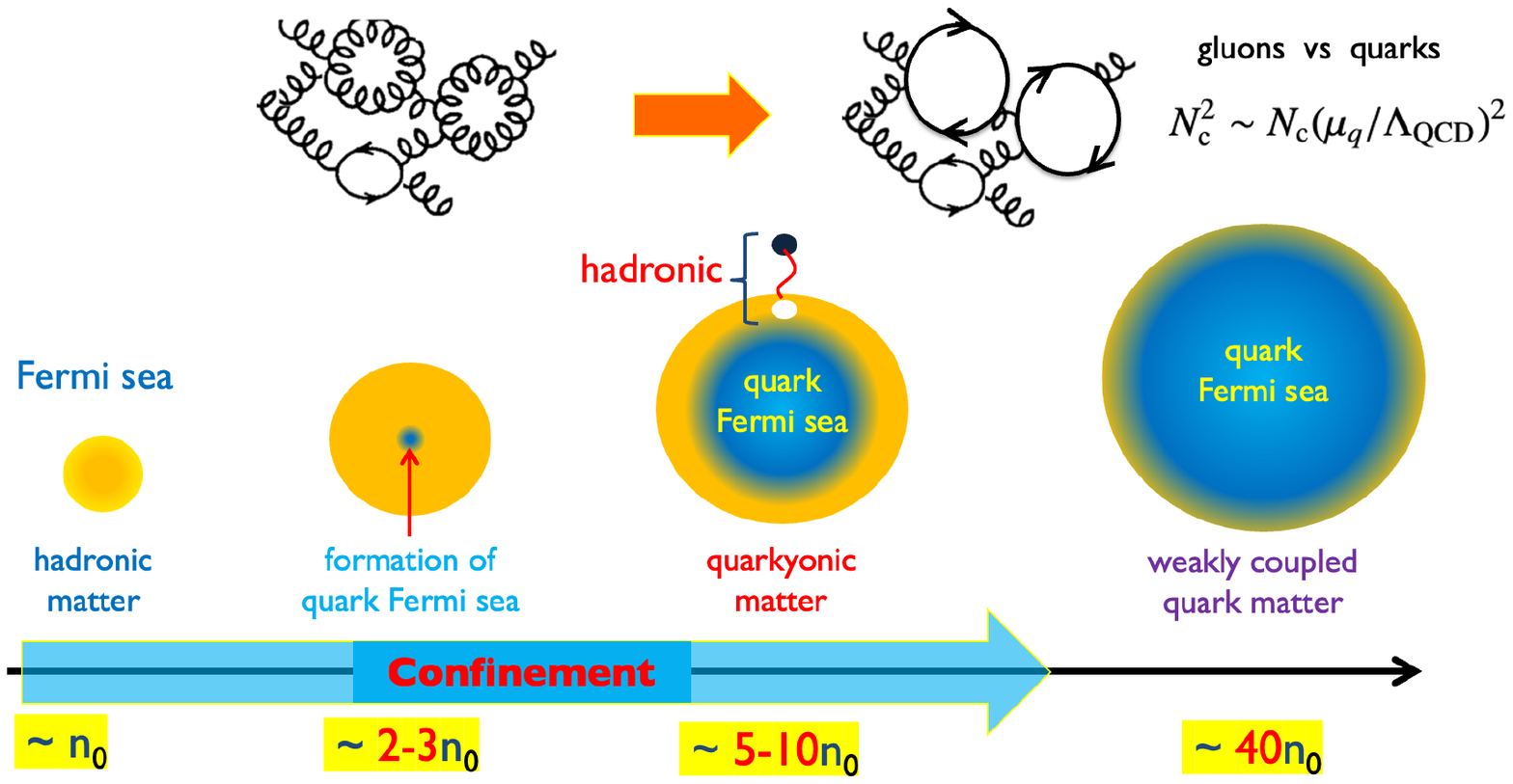}
\vspace{-1.cm}
\caption{ 
A possible scenario for the evolution of strongly interacting matter from the nuclear to the quark regime.
At large $\Nc$, gluonic contributions dominate over quark screening effects up to
$\mu_q \sim \Nc^{1/2}\Lambda_{\rm QCD}$, allowing confinement to persist at high density.
As a result, a quarkyonic regime emerges, characterized by a quark Fermi sea in the bulk
with confined, hadronic excitations near the Fermi surface.
Estimates for the real world with $\Nc=3$ are shown as a guideline for relevant densities.
%
}
\label{fig:quarkyonic}
\end{figure}

The application of large $\Nc$ leads to a new look into the relation between baryonic and quark matter.
Below we introduce the concept of {\it quarkyonic} matter \citep{McLerran:2007qj}, a hypothetical phase of QCD at large $\Nc$,
characterized by a quark Fermi sea coexisting with confining gluodynamics.

One significant difference from hot matter is that interactions among baryons are very strong.
As we discussed in Sec.~\ref{sec:baryons},
two-body forces between baryons are $\sim \Nc$.
This trend continues to three-, four-, and more-body forces.
For instance, in the meson exchange picture for three-baryon forces,
there is an extra $\Nc^{-1/2}$ suppression factor at a three-meson vertex
but there is also an extra $\Nc^{1/2}$ enhancement factor in the meson-baryon coupling (Fig.~\ref{fig:many-M_B}).
As a result, the three-baryon forces can be at the same order as two-baryon forces.
In the dilute regime, what suppresses many-baryon forces is the probability that
many-baryons meet at once;
for short-range (contact-like) interactions, the $n$-body forces scale as $V_{n-{\rm body} } \sim \Nc \lqcd \times (n_B/\lqcd^3)^n$
where $\lqcd^{-1}$ is the typical length scale for the baryon size or the range of meson-exchanges.

Once $\mu_B$ exceeds $m_N$, the baryon density increases quickly;
the baryon Fermi energy $E_F = \sqrt{ m_N^2 + p_F^2} \simeq m_N + p_F^2/2m_N$
($p_F$: baryon Fermi momentum)
changes only slightly for increasing $p_F$
so that little growth of $\mu_B$ allows $p_F$ or $n_B\sim p_F^3$ to increase rapidly.
Hence, within a small window of $\mu_B$,
the baryonic matter changes from the dilute to dense regime,
and those baryons interact strongly.
Remembering that quarks are exchanged in meson-exchange processes,
it is difficult to differentiate strongly interacting baryonic matter from quark matter.

In terms of EOS, these two descriptions of matter have interesting duality.
The discussion is simple for the energy density;
both baryonic and quark matter scale as $\varepsilon \sim \Nc \lqcd^4$ for $n_B \sim \lqcd^3$.
In baryonic descriptions, the energy density is dominated by the mass and potential energies, 
$\varepsilon_B \sim m_N n_B + \sum_n  \Nc \lqcd^4 \times (n_B/\lqcd^3)^n$,
where the kinetic term of $\sim 1/\Nc$ is neglected.
In quark descriptions, 
there are $\Nc$ quarks so that the energy scales like $\varepsilon_q \sim \Nc (E_q + V_q) n_B$
where $V_q$ is the average potential energy for a single quark.
We note that the quark density for a given color is equal to baryon density, i.e., $n_q^R = n_q^G=n_q^B =\cdots = n_B$,
since a single baryon contains one quark for a given color.

Meanwhile the pressure is more non-trivial.
The pressure can be computed as $P = n_B^2 \partial (\varepsilon/n_B)/\partial n_B$. 
For baryonic matter,
the mass energy, with the $\varepsilon_{\rm mass}/n_B \sim m_N$, 
drops off from the pressure, and the interaction terms dominate the pressure,
\beq
P_B \sim \sum_n  (n-1) \Nc \lqcd^4 \times (n_B/\lqcd^3)^n \,.
\eeq
Meanwhile, in quark descriptions,
the pressure is $P_q \sim \Nc n_B^{4/3} $ in the relativistic limit and
$P_q \sim \Nc n_B^{5/3}/M_q$ for the non-relativistic case.
At $n_B \sim \lqcd^3$,
baryonic and quark descriptions can be matched in both the energy density and pressure.
In this regard, the thermodynamic properties at large $\Nc$ may smoothly change
from the hadronic to quark matter.

The formation of quark matter, however, by itself does not imply deconfinement of colors or vanishing of color-flux tubes.
At large $\Nc$, confinement persists up to parametrically large baryon chemical
potential, $\mu_q \sim \Nc^{1/2} \lqcd$, since quark loop effects are suppressed by $1/\Nc$.
This estimate is based on the balance
between the gluonic quantum fluctuations of $\sim \Nc^2$
with the quark quantum fluctuations of $\sim \Nc (\mu_q/\lqcd)^2$ (Fig.~\ref{fig:quarkyonic}).
The enhancement factor of $(\mu_q/\lqcd)^2$ comes from the enlarged phase space near the Fermi surface, 
$\sim 4\pi p_F^2 \lqcd \sim 4\pi \mu_q^2 \lqcd $,
larger than the phase space in the vacuum cases, $\sim \lqcd^3$.
More explicitly, one can compute the Debye screening mass, $m_D \sim g_s \mu_q$, for electric gluons
and compare it with $\lqcd$.
When $\mu_q \sim \Nc^{1/2} \lqcd $, the color screening caused by quarks cuts off gluons in the infrared, 
and then the dynamics becomes weak coupling from high to low momentum transfer processes.

The above arguments suggest that, in the window between $\mu_q \sim \lqcd$ and $\mu_q \sim \Nc^{1/2} \lqcd$,
we have a matter in which natural degrees of freedom are quarks, but confining gluons persist. 
Since quarks fill the Fermi sea equally for each color, bulk quark matter can remain globally color-singlet, 
and the formation of a quark Fermi sea does not contradict confinement.
Such quark matter with confining gluons is called {\it quarkyonic} matter \citep{McLerran:2007qj}. 
In this matter, a colored-excitation on top of the color-singlet Fermi sea must be combined with other colored excitations
until the resulting composite becomes color-singlet.
Hence, it is convenient to describe quarkyonic matter as quark matter in the bulk part but baryonic matter near the Fermi surface.

An important application of the quarkyonic matter scenario is the construction of EOS \citep{McLerran:2018hbz}.
Assuming quarks confined inside baryons but increasing density,
it has been argued that quarks inevitably form the quark Fermi sea at low momenta, leading to quark scaling of the EOS \citep{Kojo:2021ugu}.
In terms of baryons, baryons under-occupy states at low momenta and largely pushed to higher momenta to avoid the breaking of the quark Pauli principle.
The onset of the quark Fermi sea triggers the rapid stiffening, as requested from neutron star constraints.
This was explicitly demonstrated for an ideal, analytically tractable model constructed in \cite{Fujimoto:2023mzy}.

In reality of $\Nc=3$, whether the window for quarkyonic matter exists or is large enough is not known
and further discussions are called for.
Especially, color-superconducting phases (for a review, \cite{Alford:2007xm}), which are plausible at high density,
are entirely missed in the large $\Nc$ limit.
Even so, the large $\Nc$ arguments have offers new perspectives on dense matter;
they certainly sharpened the conception of confinement-deconfinement and also brought our attention 
to new ideas that would explain rapid stiffening of EOS.

\section{ Two-color QCD and isospin QCD as laboratories of dense QCD } \label{sec:QC2D}

So far we have seen that the $1/\Nc$ expansion provides us with useful classification schemes from hadronic physics to extreme matter.
The next step is to examine the concepts developed in the large $\Nc$, e.g., quarkyonic matter,
by studying  theories where one can work out various computations explicitly.
Two color QCD is an example of such theories since the lattice Monte-Carlo simulations can be performed
for cold, dense matter,
without encountering the notorious sign-problem \citep{Hands:1999md,Iida:2022hyy}.
In this theory, baryonic matters are made of bosonic baryons (color-singlet diquarks) and differ from those in real-world QCD.
Nevertheless, at densities high enough for hadrons to strongly overlap,
quark matter should be formed and here the difference in colors may not be so important.
Another useful theory is isospin QCD, i.e., QCD at finite isospin but zero baryon chemical potentials.
In both QC$_2$D and QCD$_I$, dense hadronic matter is made of bosons, 
color-singlet diquarks for QC$_2$D and mesons with isospin charges (e.g., charged pions) for QCD$_I$.
Below we consider QC$_2$D unless otherwise stated, 
but for EOS we quote results in QCD$_I$ where simulations cover densities up to the pQCD domain \citep{Abbott:2023coj}.
At finite temperature and zero baryon chemical potential,
the qualitative trend is very similar to the real-world QCD;
finite temperature transitions at zero chemical potentials are crossover for quark masses which have been used for lattice simulations.
Hence below we mostly focus on the physics at finite density.

\subsection{Pseudo-real representations and Pauli--G\"ursey symmetry}
\label{sec:rep_QC2D}

To identify relevant effective degrees of freedom in QC$_2$D, it is essential to know the symmetry breaking pattern
and the associated Nambu-Goldstone (NG) bosons. 
For massless quarks, the real-world QCD has the $SU(\Nf)_L \times SU(\Nf)_R \times U(1)_B$ global symmetry.
In QC$_2$D, the symmetry is enlarged due to the pseudo-real nature of the fundamental representation of $SU(2)_c$,
which relates fundamental and anti-fundamental color representations \citep{Kogut:2000ek}.
Historically, the particle--antiparticle mixing underlying this symmetry
was first identified by Pauli \citep{Pauli1957}, while its interpretation as an internal
continuous symmetry was later formalized by G\"ursey \citep{Gursey:1958fzy}.

Using $\sigma_A$ ($A=1,2,3$) as the generators of $SU(2)_c$, we define the charge-conjugated quark field as
($C$ denotes the charge-conjugation matrix)
\beq
\big( \tilde q_{L/R} \big)^a_i
\equiv
C_{ij}\, \sigma_2^{ab}\, \big( \bar q_{R/L} \big)^b_j \, .
\eeq
By construction, $\tilde q_{L/R}$ transforms as a left-/right-handed quark
under Lorentz and flavor transformations, while transforming in the fundamental representation under color rotations.
The latter follows from
\beq
\sigma_2 q^* 
\rightarrow 
\sigma_2 U^* q^* = \sigma_2 U^* \sigma_2 \sigma_2 q^* = U \big( \sigma_2 q^* \big) \,,
~~~~~~~~~~
\sigma_2 \sigma_a^* \sigma_2 = - \sigma_a \,,
\eeq
where $U \in SU(2)_c$.
Using the identity $C \gamma_\mu C^{-1} = - \gamma_\mu^T$, the Lagrangian for massless quarks
($D_\mu = \partial_\mu + i g_s A_\mu$) can be rewritten as
\beq
\calL_q
=
\bar{q}_L \rmi \Slash{D}  q_L 
+ \bar{q}_R \rmi \Slash{D}  q_R
= \bar{q}_L \rmi \Slash{D}  q_L +  \bar{ \tilde{q} }_L \rmi \Slash{D} \tilde{q}_L
= 
\bar{\Psi}_L
\rmi \Slash{D}   
\Psi_L \,,
~~~~~~~~~~~~ 
\Psi_L \equiv ( q_L, \tilde{q}_L )^T = ( u_L, d_L, \tilde{u}_L, \tilde{d}_L )^T
\,.
\eeq
The field $\Psi_L$ transforms as a left-handed fermion under Lorentz transformations.
Alternatively, one may work with $\Psi_R$ instead of $\Psi_L$.

Under the usual chiral rotations
$SU(\Nf)_L \times SU(\Nf)_R$, the fields transform independently as
$q_L \rightarrow e^{i \theta_L^f \tau_f} q_L$
and
$\tilde q_L \rightarrow e^{i \theta_R^f \tau_f} \tilde q_L$,
leaving the Lagrangian invariant.
In addition, the Lagrangian is invariant under the enlarged
$SU(2 \Nf)$ transformation
$\Psi_L \rightarrow e^{i \theta^f T_f} \Psi_L$,
where $T_f$ are the generators of $SU(2 \Nf)$.
This enlarged $SU(2N_f)$ symmetry, which mixes quarks and charge-conjugated
antiquarks, is known as the Pauli--G\"ursey symmetry
\citep{Pauli1957,Gursey:1958fzy}, and strongly constrains
the dynamics of mesons and diquarks.
We note that the $U(1)_B$ symmetry, which acts differently on quarks and anti-quarks,
is already embedded in $SU(2N_f)$.
Therefore, in the massless case, the number of generators is $(2N_f)^2 - 1$,
which equals 15 for $N_f=2$.

Upon introducing the quark mass term, it is tempting to employ both $\Psi_L$ and $\Psi_R$. 
But their chiral transformations are not independent by construction
and hence there is no
$SU(2N_f)_L \times SU(2N_f)_R$ symmetry even in the massless limit.
Using $\Psi_L$ fields alone, the mass term takes the form
\beq
 \bar{q}_L \hat{m_q} q_R
 +  \bar{q}_R \hat{m_q} q_L
= 
\Psi^T_L C \sigma_2 
\begin{bmatrix}
0 & - \hat{m}_q \\
\hat{m}_q & 0 
\end{bmatrix}
 \Psi_L 
\,,~~~~~
\eeq
where $\hat m_q$ denotes the quark mass matrix.

From here on, we restrict ourselves to the $N_f=2$ case and assume degenerate
current quark masses, $m_u = m_d$.
The mass term explicitly breaks the global flavor $SU(4)$ symmetry 
to $Sp(4) \simeq SO(5)$,
which is defined as the subgroup
of unitary transformations satisfying
\beq
U^T \Sigma_0 U = \Sigma_0\,,
~~~~~~~~
\Sigma_0 
=
\begin{bmatrix}
0 & {\bf 1}_2 \\
- {\bf 1}_2 & 0 
\end{bmatrix} 
\,.
\eeq
Likewise, the formation of the chiral condensate $\la \bar{u} u \ra = \la \bar{d} d \ra$
spontaneously breaks the symmetry as
$SU(4) \rightarrow Sp(4)$.
Since the number of broken generators is $15 - 10 = 5$,
five NG bosons emerge.
These NG bosons form a five-dimensional multiplet of $SO(5)$ with a common mass $m_\pi$.
They consist of three pions, together with a Lorentz-scalar,
flavor-singlet diquark and anti-diquark.

\subsection{ Phase structure } \label{sec:phase_QCDlike}

\begin{figure}[t]
\vspace{-1.cm}
\centering
\includegraphics[width=.5\textwidth]{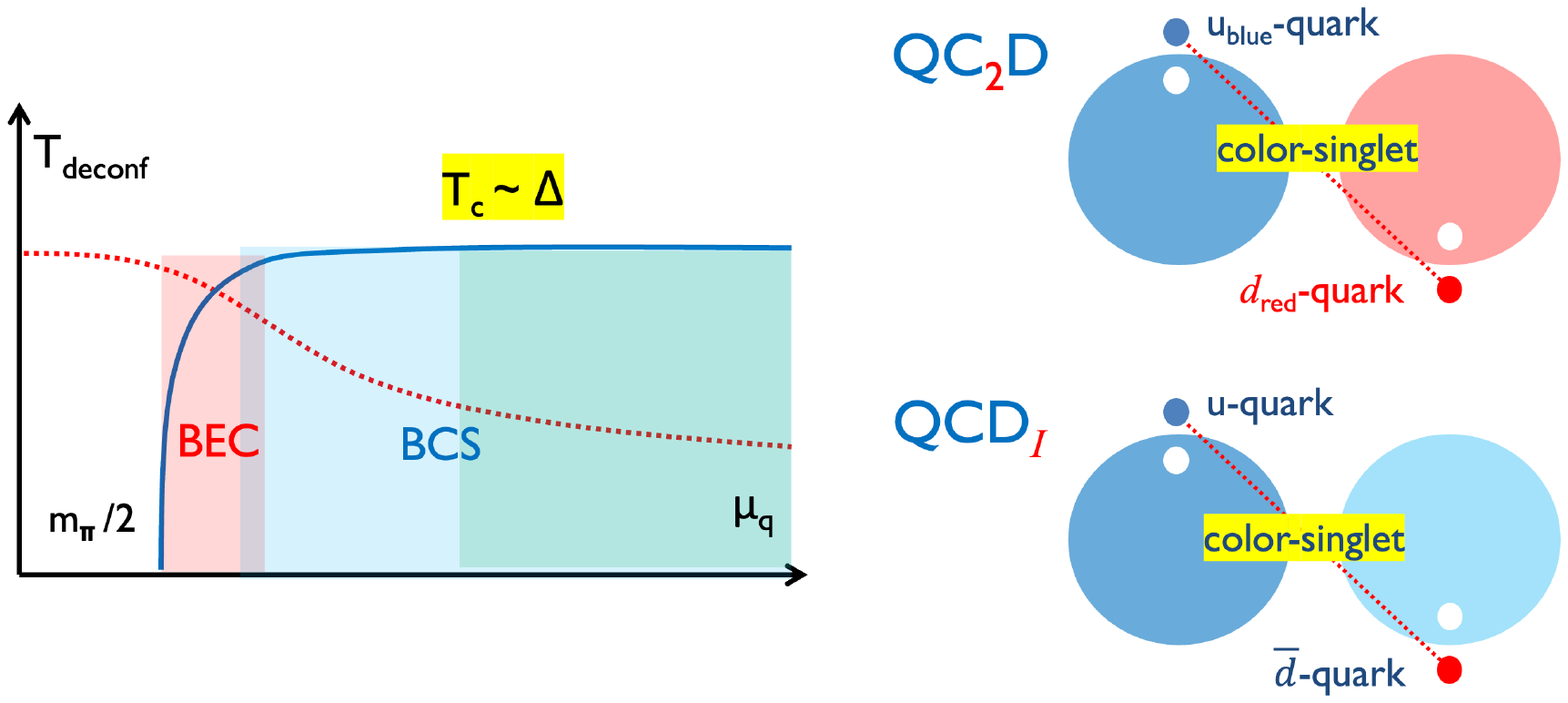}
\vspace{-1.cm}
\caption{ 
The phase diagram for QC$_2$D (QCD$_I$).
Color singlet diquarks (charged pions $\pi_+$) begin to condense at $\mu = \mu_q = m_\pi/2$ ($\mu = \mu_I/2$).
The dense matter is created first in the BEC regime and crossover to the BCS regime.
}
\label{fig:qcd2_qcdI_phase}
\end{figure}

At finite baryon density, the qualitative features of hadronic matter
are substantially modified.
In particular, the presence of a nonzero baryon chemical potential $\mu_B$
explicitly breaks the $Sp(4)$ symmetry down to
$SU(2)\times U(1)_B$,
since $\mu_B$ differentiates quarks and anti-quarks.
Below we assume the current quark mass to be nonzero
so that the chiral symmetry is explicitly broken.

Shown in Fig.~\ref{fig:qcd2_qcdI_phase} is the phase diagram for QC$_2$D (which can be also used for QCD$_I$ after slight modifications).
When the baryon chemical potential reaches the diquark mass,
$\mu_B = m_\pi$ (or equivalently $\mu_q = m_\pi/2$),
diquarks begin to form Bose--Einstein condensates,
thereby spontaneously breaking the $U(1)_B$ symmetry.
The condensed diquarks are composed of $u$- and $d$-quarks in the
spatial $s$-wave, spin-singlet, and flavor-singlet channel.

As the baryon density increases,
diquarks eventually overlap and quarks inside those diquarks become important.
Those quarks form the Fermi sea
and the system is more naturally interpreted as quark matter 
rather than hadronic matter.
Similarly, the diquark condensation
changes from the BEC to the Bardeen-Copper-Schrieffer (BCS) regime (see e.g., \cite{Sun:2007fc,vonSmekal:2012vx}).
This transition is a BEC-BCS crossover, with no change in symmetry and no discontinuity in thermodynamic quantities.

These low temperature phases with the diquark condensates
can be clearly distinguished from the high temperature phases without the diquark condensates,
since $U(1)_B$ symmetry is broken in the former but not in the latter.
This situation differs from chiral transitions
where a small but explicit breaking due to the quark mass exists
and the spontaneous symmetry breaking is only an approximate concept.

Although the BEC-BCS crossover does not involve any symmetry change,
the relevant degrees of freedom gradually shift from hadronic
to quark-like ones.
This change becomes particularly important for the equation of state,
where Pauli blocking and quark occupation effects play a crucial role.

\subsection{ Color-singlet quark gap, screening, and gluons } \label{sec:gluons}

We consider the mean-field spectra of quarks in the BCS regime
and examine the impacts on the low energy gluon dynamics.
We assume the Yukawa coupling between quarks and (condensed) diquarks (${\rm d_q}$),
\beq
\frac{\, g \,}{2} \la {\rm d_q} \ra \epsilon_{ij} \epsilon_{ab} \big[ ( \bar{q}_C )_i^a  \rmi \gamma_5 q_j^b  \big] + {\rm h.c.}
= - \frac{\, g \,}{2} \la {\rm d_q} \ra \big[ \bar{q}_C \sigma_2 \tau_2  \rmi \gamma_5 q \big] + {\rm h.c.}
= \bar{\psi} 
\begin{bmatrix}
0 & \rmi \gamma_5 \Delta \\
- \rmi \gamma_5 \Delta & 0 
\end{bmatrix} 
\psi \,,
~~~~~~
\psi 
= 
\frac{ 1 }{\, \sqrt{2} \,} \left(
\begin{matrix}
q \\
\sigma_2 \tau_2 q_C
\end{matrix}
\right) \,,
\eeq
where $q_C = C \bar{q}^T$ and $\Delta = g \la {\rm d_q} \ra$.
Here $\psi$ is called the Nambu-Gor'kov spinor.
The mean-field effective Lagrangian is
\beq
\calL_{\rm MF}
= 
 \bar{\psi} (p)
\begin{bmatrix}
\Slash{p} + \mu \gamma_0 - m_q & \rmi \gamma_5 \Delta \\
- \rmi \gamma_5 \Delta & \Slash{p} - \mu \gamma_0 - m_q
\end{bmatrix} 
\psi (p) \,.
\eeq
Diagonalizing the matrix, one finds the spectra
$\epsilon_{\rm p/a} (\vp) = \sqrt{ (E_{\vp} \mp \mu_q)^2 + \Delta^2 }$
with $E_{\vp} = \sqrt{ \vp^2 + m_q^2 }$.
The gap $\Delta$ is the minimum energy to place a quark on top of the quark Fermi sea.

This gap for quark excitations has important impacts on the Debye (electric) screening in colors.
The Debye screening mass for gluons
can be estimated by computing quark loops coupled to gluon propagators
(see, e.g., \cite{Rischke:2000ra,Rischke:2000qz}).
For $\Delta =0$, the polarization loop at low momentum $\vk$ is
dominated by particle-hole excitations,
\beq
\Pi_{\rm ph} (\vk)
\sim
g_s^2 \int_{\vp}  \, 
\frac{\, \Theta \big( E_{\vp+\vk} -\mu_q \big) \Theta \big( \mu_q - E_{\vp} \big) 
+ \Theta \big( \mu_q - E_{\vp+\vk}  \big) \Theta \big( E_{\vp} - \mu_q \big) \,}{\, | E_{\vp+\vk} -\mu_q | + | E_{\vp} - \mu_q | \,}
\sim
g_s^2 \int_{\vp}  \, 
\frac{\,  \vec{v} \cdot \vk \, \delta \big( E_{\vp} - \mu_q \big) \,}{\, E_{\vp+\vk} - E_{\vp}  \,} 
\sim
g_s^2 p_F E_F + O(\vk^2)\,.
\label{eq:pol}
\eeq
Thus, in the static and long-wavelength limit,
$\Pi_{\rm ph} (\vk)$ approaches a constant Debye mass,
$m_D^2 \sim ( g_s \mu_q )^2$.
The key observation is that, in the integrand, 
both the numerator and denominator are vanishing for $\vk\rightarrow 0$, but the ratio remains finite.

In contrast, in the presence of a quark gap $\Delta$,
the particle-hole excitation energy is bounded from below by
$\varepsilon_{\rm p} (\vp+\vk) + \varepsilon_{\rm p} (\vp) \ge 2\Delta$,
i.e.,  the energy denominator in Eq.~\eqref{eq:pol} becomes finite at $\vk \rightarrow 0$.
As a result, the static polarization vanishes in the infrared,
and the Debye screening disappears.
Hence, the gluon propagators $D_g(k)$ in the infrared, with $|\vk| \le \Delta$, are protected from the in-medium screening effects.
This can leave the non-perturbative gluons even in high density quark matter,
a system as discussed in the quarkyonic matter scenario.
The lattice simulations indeed seem to be consistent with this observation \citep{Boz:2018crd}.

The presence of the gap by itself does not mean the absence of the color screening.
What is special for QC$_2$D is that the condensate is color-singlet,
which guarantees that the numerator, characterized by the quark wavefunctions,
vanishes in the limit $\vk \rightarrow 0$ (see, e.g., \cite{Kojo:2014vja}).
If the diquark condensate is colored, as in the color-superconductivity,
both the numerator and denominator are nonzero, leaving the Debye mass of $\sim g_s \mu_q$.

As for the magnetic screening, the screening mass vanishes as far as the condensate does not have colors.
Hence in QC$_2$D magnetic gluons in the infrared are also protected (as in the normal phase).
This vanishing is achieved by mutual cancellations among the particle-hole and particle-antiparticle contributions
which are constrained by the gauge invariance (see, e.g., \cite{Kojo:2014vja}).
In contrast, in the color-superconducting phase in QCD, this cancellation does not hold and the magnetic (Meissner) mass is generated. 

Concerning diquark condensed phases, the gluon dynamics in QC$_2$D cannot be immediately transferred
to real-world QCD, since the color structure of the condensates is fundamentally different.
Whether gluons in dense QCD remain non-perturbative is therefore a quantitative question,
and further studies are needed to estimate the density window
where non-perturbative gluon dynamics may persist.

\subsection{ Equations of state } \label{sec:EOS_QC2D}

\begin{figure}[t]
\vspace{-2.cm}
\centering
\includegraphics[width=.65\textwidth]{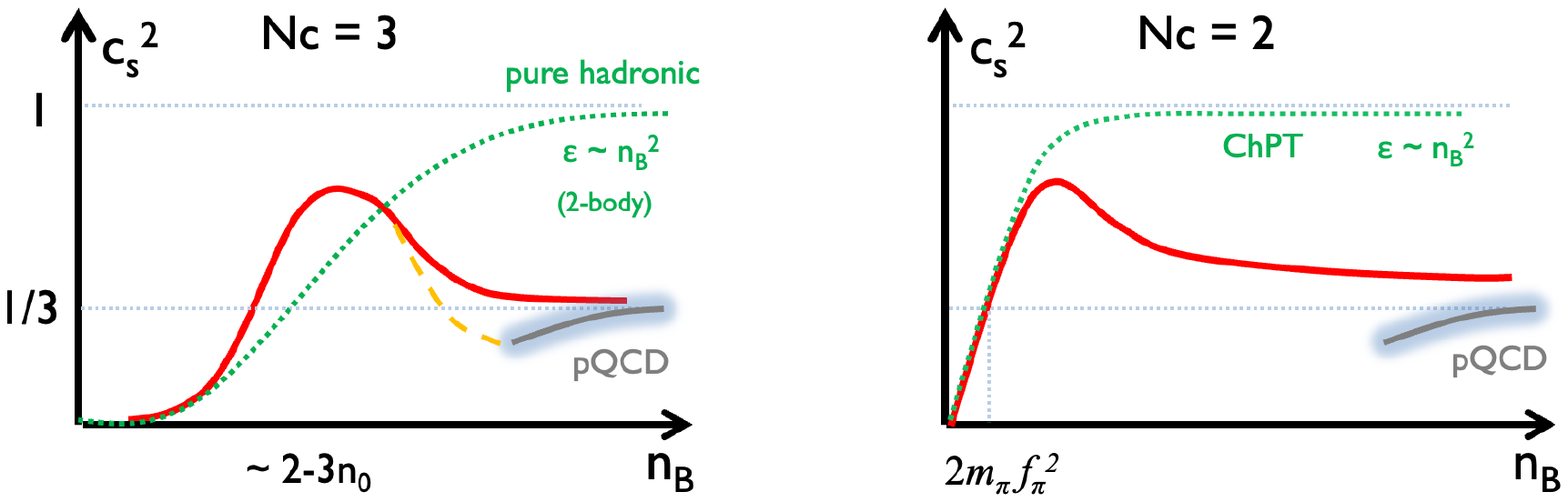}
\vspace{-2.cm}
\caption{ 
The density evolution of squared sound speed $c_s^2$ for three- and two-color QCD.
(i) For three colors, a matter made of non-relativistic baryons is soft at low density
but rapidly becomes stiff around $n_B\sim$ 2-3$n_0$.
How $c_s^2$ approaches the pQCD estimate at high density is not well-known.
(ii) For two colors, EOS of light diquarks become quickly dominated by interactions.
}
\label{fig:cs2_2color}
\end{figure}

With the rapid progress in neutron star observations,
QC$_2$D has attracted renewed interest in the context of the stiffening of EOS \citep{Kojo:2021hqh}.
As discussed in Sec.~\ref{sec:EOS},
neutron star data suggest the presence of a peak in the sound speed squared
exceeding the conformal limit $c_s^2 = 1/3$.
While several phenomenological models had indicated such behavior,
it remained unclear whether these results reflected genuine physics
or artifacts of specific models and uncontrolled extrapolations
beyond the domain of applicability of low density EOS.
The first reliable evidence for a sound-speed peak came from lattice simulations
of QC$_2$D performed by Iida and Itou \citep{Iida:2022hyy}.
Similar behavior was also confirmed in QCD$_I$ \citep{Brandt:2022hwy,Abbott:2023coj}.
Qualitative comparisons of $c_s^2$ for QCD and QC$_2$D (QCD$_I$) are shown in Fig.~\ref{fig:cs2_2color}.

\subsubsection{ Dilute regime } \label{sec:dilute_QC2D}

In the dilute regime near $\mu_B \sim m_\pi$, the EOS is well-described by the chiral perturbation theory (ChPT),
a low-energy effective field theory of QCD based on the spontaneous breaking of chiral symmetry \citep{Weinberg1979,GasserLeutwyler1984,GasserLeutwyler1985}.
Computing $P(\mu_B)$ first, subsequently the number and energy densities can be computed as \citep{Kogut:2000ek}
\beq
P(\mu_B) 
= \frac{\, f_\pi^2 \,}{2} \mu_B^2 \left( 1 - \frac{\, m_\pi^2 \,}{\, \mu_B^2 \,} \right)^2
~~~~~
\rightarrow ~~~~ 
n_B = \frac{\, \partial P \,}{\, \partial \mu_B \,}
= f_\pi^2  \mu_B \left( 1 - \frac{\, m_\pi^4 \,}{\, \mu_B^4 \,} \right)\,,
~~~~~~ 
\varepsilon = \mu_B n_B - P 
= \frac{\, f_\pi^2 \,}{2}  \left( \mu_B^2 + 2m_\pi^2  - \frac{\, 3 m_\pi^4 \,}{\, \mu_B^2 \,} \right)
\,.
\eeq
A more intuitive form can be derived by rewriting $\varepsilon$ as a function of $n_B$.
Near the onset, we expand $\mu_B \simeq m_\pi + \delta \mu_B$,
express $n_B \simeq 4 f_\pi^2 \delta \mu_B + O(\delta \mu_B^2)$,
and then eliminate $\delta \mu_B$ of $P$ in favor of $n_B$.
In the dilute regime, the resulting expression consists of the mass energy and two-body repulsive terms,
\beq
\varepsilon \simeq m_\pi n_B + \frac{\, n_B^2 \,}{\, 8 f_\pi^2 \,} + O(n_B^3) \,.
\eeq 
%
Unlike the real-world QCD, (i) the baryon is so light that the non-relativistic regime ends within a small density interval;
(ii) there is no kinetic term like $\sim n_B^{5/3}/m_\pi$ since,  in bosonic systems, there is no Fermi momentum for baryons.

The mass and interaction terms become comparable when $n_B \sim 2 m_\pi f_\pi^2$ 
beyond which the validity of the above simple expression would be lost.
If we assume $m_\pi = 140$ MeV and $f_\pi = 90$ MeV, the density for the breakdown is $2 m_\pi f_\pi^2 \simeq 0.30\,{\rm fm}^{-3}$,
which is about a half of the density where diquarks (pions) with the radii $\sim 0.7$ fm to spatially overlap.
The pressure as a function of $n_B$ and the sound speed is computed as
\beq
P \simeq \frac{ n_B^2 }{\, 8 f_\pi^2 \,} \,,~~~~~~~~
c_s^2 = \frac{\, \rmd P/\rmd n_B \,}{\, \rmd \varepsilon/\rmd n_B \,} \simeq \frac{\, n_B \,}{\, 4 m_\pi f_\pi^2 + n_B \,} \,.
\label{eq:cs2_ChPT}
\eeq
The $c_s^2$ increases with $n_B$ from zero, exceeds the conformal limit $1/3$ at $n_B \simeq 2 m_\pi f_\pi^2$, 
and asymptotically reaches the causal upperbound $c_s^2 =1$ (the sound speed should not exceed the speed of light; see, e.g., \citep{Zeldovich1962}).
But $c_s^2 > 1/3$ is achieved only beyond the supposed breakdown scale $n_B \sim 2 m_\pi f_\pi^2$ 
and hence the validity of Eq.~\eqref{eq:cs2_ChPT} is questionable.
In particular, the limit $c_s^2 \rightarrow 1$ at large $n_B$ is a mere artifact 
of extrapolating the dilute chiral expansion beyond its domain of validity.
It is worth mentioning that, if the EOS were dominated by an $N$-body interaction
term of the form $\varepsilon \sim n_B^N$, one would formally obtain the asymptotic
behavior $c_s^2 \sim N-1$, which exceeds the causal bound $c_s^2=1$ for $N>2$ (see, e.g., \cite{Kojo:2025vcq}).
This observation illustrates that a naive power-series expansion in $n_B$,
when extrapolated to high density, is incompatible with basic relativistic constraints.
Therefore, the structure of the EOS must be qualitatively modified in the dense regime,
where new degrees of freedom and nonperturbative dynamics become essential.

\subsubsection{ Dense regime } \label{sec:dense_QC2D}

In the dense regime after baryons overlap, one expects the importance of quark degrees of freedom.
In the relativistic limit,
the kinetic energy is so large that it dominates over the interaction energy.
The EOS behaves as $P(\mu_q) \sim \mu_q^4$ or, equivalently, $\varepsilon (n_B) \sim n_B^{4/3}$.

At high density $\mu_q > 1$-2 GeV, one expects the applicability of 
perturbative QCD (pQCD) \citep{Freedman:1976ub,Kurkela:2009gj,Gorda:2023mkk}.
The pressure of perturbative QCD at asymptotically high density
can be written as
\beq
P(\mu_q)
= \frac{\, \Nc \Nf \,}{\, 12\pi^2 \,}\,\mu_q^4
\left[
1
- a_2\,\alpha_s(\mu_q)
- a_4 \,\alpha_s^2(\mu_q)
- a_4' \,\alpha_s^2(\mu_q)\ln\alpha_s(\mu_q)
+ \cdots
\right],
\eeq
where $a_i>0$ are numerical coefficients depending on $\Nf$
and the renormalization scale for $\alpha_s$ is taken to be $\sim \mu_q$.
The logarithm $\ln \alpha_s$ comes from the ratio between $\mu_q$ and the Debye mass $m_D \sim g_s \mu_q$.
In pQCD, the intrinsic scale of QCD, $\lqcd$, appears only through $\alpha_s$.
From this expression one can compute $c_s^2$ which, as $\mu_q$ increases, approaches the conformal limit $1/3$ from below.
Correspondingly, pQCD suggests $c_s^2 < 1/3$ in the density range 
where perturbation theory is expected to become applicable, $\mu_q \sim 1$-2 GeV.

How $c_s^2< 1/3$ at high density can be consistent with $c_s^2 > 1/3$ at low density is an important question
in contemporary dense QCD studies.
One possible scenario is to add (non-perturbative) power corrections,
of the form $\sim \mu_q^2 \lqcd^2$, to the pQCD EOS.
Since $\lqcd$ is non-analytic as a function of $\alpha_s$,
the expansion of $\alpha_s$ cannot capture such corrections.
A simple EOS with the power corrections and the resultant $c_s^2$ are
(e.g., \cite{Chiba:2023ftg})
\beq
P(\mu_q) = c_0 \mu_q^4 + c_2 \mu_q^2 \lqcd^2
~~~\rightarrow~~~
c_s^2 = \frac{\, 2c_0 \mu_q^2 + c_2 \lqcd^2 \,}{\, 6c_0 \mu_q^2 + c_2 \lqcd^2 \,} \,.
\eeq
Assuming $c_2 >0$, we find $c_s^2 > 1/3$ which approaches the conformal limit from above.
Increasing density in the BEC regime and reducing density in the BCS regime
both lead to the sound speed peak in the crossover domain.

How the power corrections arise and what physics is responsible for $\lqcd$ depend on theories.
One plausible origin for $\lqcd$ is the gap $\Delta$ in diquark condensed phases.
In the BCS theory the presence of the gap adds a term like $\delta P \sim + \mu_q^2 \Delta^2$ to the pressure
or more intuitively leads to the energy reduction, $\delta \varepsilon \sim - p_F^2 \Delta^2 $,
with $p_F^2$ comes from the area of the Fermi surface, $\sim 4 \pi p_F^2$.

If $\Delta$ depends on $\mu_q$ only weakly and is comparable to $\lqcd$,
the BCS-type EOS immediately supplies the power corrections $\mu_q^2 \lqcd^2$, 
smoothly matching the low density EOS having the $c_s^2$ peak with the high density counterpart.
This picture is supported for the case of isospin QCD (QCD$_I$),
where lattice simulations up to the pQCD domain are available \citep{Abbott:2023coj}. 
The data shows that $c_s^2$ has a peak at low density and remains larger than $1/3$ at $\mu_q \sim 1$ GeV.
The data is well-reproduced for a nearly constant $\Delta$ of $\sim 200$-$300$ MeV.

To derive the gap $\Delta$ at high density, it is natural to apply weak coupling approaches \citep{Son:1998uk}.
Several analyses showed that the overall size and density dependence of $\Delta$ 
are sensitive to theories we consider;
$\Delta$ for the pion condensed phase in QCD$_I$, diquark condensed phase in QC$_2$D, 
and color-superconducting phase in the real-world QCD, can be considerably different,
since different numerical factors appear in the exponent \citep{Wang:2001aq,Fujimoto:2023mvc}.
For QCD$_I$, the weak coupling estimate is consistent with the lattice data,
while in QC$_2$D the estimate seems smaller by a factor of $\sim 2$-4.
The prediction for two-flavor pairing in QCD is $\Delta $ of a several MeV, too small for power corrections to alter 
the trend of pQCD.
In the real-world QCD,
how to reconcile the $c_s^2$ peak inferred from neutron star data
and the weak coupling results at very high density 
remains an important question \citep{Fukushima:2024gmp}.

\section{ Summary } \label{ sec:summary }

In this review we have surveyed various aspects of QCD and QCD-like theories
from the viewpoint of the $1/N_c$ expansion, covering its applications to
hadron physics, nuclear interactions, hot QCD, and dense matter.
Despite $\Nc=3$, the $1/N_c$ expansion provides a powerful organizing
principle that reveals common structures across seemingly different regimes.

Dense QCD stands out as a particularly challenging arena for the $1/N_c$
approach, due to the sign problem, the limited experimental access, and the
restricted applicability of effective theories.
At the same time, it is precisely in this regime that the interplay between
perturbative and nonperturbative physics becomes most transparent.

In this context, QCD with two colors plays a special role.
Because baryons are bosonic and lattice simulations at finite density are
feasible, QC$_2$D provides a theoretically controlled laboratory to study dense
QCD matter from the dilute regime to asymptotically high density.
The EOS of QC$_2$D allows for a direct comparison between
ChPT, lattice results, and perturbative expectations.

A particularly illuminating observable is the sound speed.
Lattice simulations of QC$_2$D show that the sound speed exceeds the conformal
limit at intermediate densities, a feature that cannot be explained by
perturbative QCD alone.
Interestingly, neutron star observations suggest a similar behavior in real
QCD, indicating that the emergence of a sound-speed peak may be a generic
feature of dense QCD matter, largely independent of the number of colors.

These findings highlight the importance of intermediate-density physics,
where a rearrangement of degrees of freedom and nonperturbative effects become
essential.
Comparing theories with different numbers of colors thus provides valuable
insight into the structure of dense QCD and helps to bridge the gap between
controlled theoretical limits and real-world strongly correlated matter.


\begin{ack}[Acknowledgments] \\
This work was supported by JSPS KAK- ENHI Grant No. 23K03377.
 \end{ack}

\seealso{article title article title}

\bibliographystyle{Harvard}
\bibliography{ref}

\end{document}